\errorstopmode
\input amssym.def
\input amssym.tex


\magnification=\magstephalf
\hsize=14.0 true cm
\vsize=19 true cm
\hoffset=1.0 true cm
\voffset=2.0 true cm

\abovedisplayskip=12pt plus 3pt minus 3pt
\belowdisplayskip=12pt plus 3pt minus 3pt
\parindent=1.0em


\font\sixrm=cmr6
\font\eightrm=cmr8
\font\ninerm=cmr9

\font\sixi=cmmi6
\font\eighti=cmmi8
\font\ninei=cmmi9

\font\sixsy=cmsy6
\font\eightsy=cmsy8
\font\ninesy=cmsy9

\font\sixbf=cmbx6
\font\eightbf=cmbx8
\font\ninebf=cmbx9

\font\eightit=cmti8
\font\nineit=cmti9

\font\eightsl=cmsl8
\font\ninesl=cmsl9

\font\sixss=cmss8 at 8 true pt
\font\sevenss=cmss9 at 9 true pt
\font\eightss=cmss8
\font\niness=cmss9
\font\tenss=cmss10

\font\sixmib=cmmib6
\font\sevenmib=cmmib7
\font\eightmib=cmmib8
\font\ninemib=cmmib9
\font\tenmib=cmmib10

 at 12 true pt
 at 12 true pt
\font\bigrm=cmr10 at 12 true pt
 at 12 true pt
 at 12 true pt

 at 16 true pt
 at 16 true pt
\font\Bigrm=cmr12 at 16 true pt
 at 16 true pt
 at 16 true pt

\catcode`@=11
\newfam\ssfam
\newfam\mibfam

\def\tenpoint{\def\rm{\fam0\tenrm}%
    \textfont0=\tenrm \scriptfont0=\sevenrm \scriptscriptfont0=\fiverm
    \textfont1=\teni  \scriptfont1=\seveni  \scriptscriptfont1=\fivei
    \textfont2=\tensy \scriptfont2=\sevensy \scriptscriptfont2=\fivesy
    \textfont3=\tenex \scriptfont3=\tenex   \scriptscriptfont3=\tenex
    \textfont\itfam=\tenit                  \def\it{\fam\itfam\tenit}%
    \textfont\slfam=\tensl                  \def\sl{\fam\slfam\tensl}%
    \textfont\bffam=\tenbf \scriptfont\bffam=\sevenbf
                           \scriptscriptfont\bffam=\fivebf
                           \def\bf{\fam\bffam\tenbf}%
    \textfont\ssfam=\tenss \scriptfont\ssfam=\sevenss
                           \scriptscriptfont\ssfam=\sevenss
                           \def\ss{\fam\ssfam\tenss}%
    \textfont\mibfam=\tenmib \scriptfont\mibfam=\sevenmib
                             \scriptscriptfont\mibfam=\sevenmib
                             \def\mib{\fam\mibfam\tenmib}%
    \normalbaselineskip=13pt
    \setbox\strutbox=\hbox{\vrule height8.5pt depth3.5pt width0pt}%
    \let\big=\tenbig
    \normalbaselines\rm}

\def\ninepoint{\def\rm{\fam0\ninerm}%
    \textfont0=\ninerm      \scriptfont0=\sixrm
                            \scriptscriptfont0=\fiverm
    \textfont1=\ninei       \scriptfont1=\sixi
                            \scriptscriptfont1=\fivei
    \textfont2=\ninesy      \scriptfont2=\sixsy
                            \scriptscriptfont2=\fivesy
    \textfont3=\tenex       \scriptfont3=\tenex
                            \scriptscriptfont3=\tenex
    \textfont\itfam=\nineit \def\it{\fam\itfam\nineit}%
    \textfont\slfam=\ninesl \def\sl{\fam\slfam\ninesl}%
    \textfont\bffam=\ninebf \scriptfont\bffam=\sixbf
                            \scriptscriptfont\bffam=\fivebf
                            \def\bf{\fam\bffam\ninebf}%
    \textfont\ssfam=\niness \scriptfont\ssfam=\sixss
                            \scriptscriptfont\ssfam=\sixss
                            \def\ss{\fam\ssfam\niness}%
    \textfont\mibfam=\ninemib \scriptfont\mibfam=\sixmib
                            \scriptscriptfont\mibfam=\sixmib
                            \def\mib{\fam\mibfam\ninemib}%
    \normalbaselineskip=12pt
    \setbox\strutbox=\hbox{\vrule height8.0pt depth3.0pt width0pt}%
    \let\big=\ninebig
    \normalbaselines\rm}

\def\eightpoint{\def\rm{\fam0\eightrm}%
    \textfont0=\eightrm      \scriptfont0=\sixrm
                             \scriptscriptfont0=\fiverm
    \textfont1=\eighti       \scriptfont1=\sixi
                             \scriptscriptfont1=\fivei
    \textfont2=\eightsy      \scriptfont2=\sixsy
                             \scriptscriptfont2=\fivesy
    \textfont3=\tenex        \scriptfont3=\tenex
                             \scriptscriptfont3=\tenex
    \textfont\itfam=\eightit \def\it{\fam\itfam\eightit}%
    \textfont\slfam=\eightsl \def\sl{\fam\slfam\eightsl}%
    \textfont\bffam=\eightbf \scriptfont\bffam=\sixbf
                             \scriptscriptfont\bffam=\fivebf
                             \def\bf{\fam\bffam\eightbf}%
    \textfont\ssfam=\eightss \scriptfont\ssfam=\sixss
                             \scriptscriptfont\ssfam=\sixss
                             \def\ss{\fam\ssfam\eightss}%
    \textfont\mibfam=\eightmib \scriptfont\mibfam=\sixmib
                             \scriptscriptfont\mibfam=\sixmib
                             \def\mib{\fam\mibfam\eightmib}%
    \normalbaselineskip=10pt
    \setbox\strutbox=\hbox{\vrule height7.0pt depth2.0pt width0pt}%
    \let\big=\eightbig
    \normalbaselines\rm}

\def\tenbig#1{{\hbox{$\left#1\vbox to8.5pt{}\right.\n@space$}}}
\def\ninebig#1{{\hbox{$\textfont0=\tenrm\textfont2=\tensy
                       \left#1\vbox to7.25pt{}\right.\n@space$}}}
\def\eightbig#1{{\hbox{$\textfont0=\ninerm\textfont2=\ninesy
                       \left#1\vbox to6.5pt{}\right.\n@space$}}}

\font\sectionfont=cmbx10
\font\subsectionfont=cmti10

\def\figurecaptionfont{\ninepoint}
\def\tablecaptionfont{\ninepoint}


\newcount\equationno
\newcount\bibitemno
\newcount\figureno
\newcount\tableno

\equationno=0
\bibitemno=0
\figureno=0
\tableno=0


\footline={\ifnum\pageno=0{\hfil}\else
{\hss\rm\the\pageno\hss}\fi}


\def\section #1. #2 \par
{\vskip0pt plus .10\vsize\penalty-100 \vskip0pt plus-.10\vsize
\vskip 1.6 true cm plus 0.2 true cm minus 0.2 true cm
\global\def\equationlabel{#1}
\global\equationno=0
\leftline{\sectionfont #1. #2}\par
\immediate\write\terminal{Section #1. #2}
\vskip 0.7 true cm plus 0.1 true cm minus 0.1 true cm
\noindent}


\def\subsection #1 \par
{\vskip0pt plus 0.8 true cm\penalty-50 \vskip0pt plus-0.8 true cm
\vskip2.5ex plus 0.1ex minus 0.1ex
\leftline{\subsectionfont #1}\par
\immediate\write\terminal{Subsection #1}
\vskip1.0ex plus 0.1ex minus 0.1ex
\noindent}


\def\appendix #1. #2 \par
{\vskip0pt plus .10\vsize\penalty-100 \vskip0pt plus-.10\vsize
\vskip 1.6 true cm plus 0.2 true cm minus 0.2 true cm
\global\def\equationlabel{\hbox{\rm#1}}
\global\equationno=0
\leftline{\sectionfont Appendix #1. #2}\par
\immediate\write\terminal{Appendix #1. #2}
\vskip 0.7 true cm plus 0.1 true cm minus 0.1 true cm
\noindent}



\def\equation#1{$$\displaylines{\qquad #1}$$}
\def\enum{\global\advance\equationno by 1
\hfill\llap{{\rm(\equationlabel.\the\equationno)}}}
\def\noenum{\hfill}

\def\nexteq#1{\cr\noalign{\vskip#1}\qquad}


\def\ifundefined#1{\expandafter\ifx\csname#1\endcsname\relax}

\def\ref#1{\ifundefined{#1}?\immediate\write\terminal{unknown reference
on page \the\pageno}\else\csname#1\endcsname\fi}

\newwrite\terminal
\newwrite\bibitemlist

\def\bibitem#1#2\par{\global\advance\bibitemno by 1
\immediate\write\bibitemlist{\string\def
\expandafter\string\csname#1\endcsname
{\the\bibitemno}}
\item{[\the\bibitemno]}#2\par}

\def\beginbibliography{
\vskip0pt plus .15\vsize\penalty-100 \vskip0pt plus-.15\vsize
\vskip 1.2 true cm plus 0.2 true cm minus 0.2 true cm
\leftline{\sectionfont References}\par
\immediate\write\terminal{References}
\immediate\openout\bibitemlist=biblist
\frenchspacing\parindent=1.8em
\vskip 0.5 true cm plus 0.1 true cm minus 0.1 true cm}

\def\endbibliography{
\immediate\closeout\bibitemlist
\nonfrenchspacing\parindent=1.0em}


\def\figurecaption#1{\global\advance\figureno by 1
\narrower\figurecaptionfont
Fig.~\the\figureno. #1}

\def\tablecaption#1{\global\advance\tableno by 1
\vbox to 0.25 true cm { }
\centerline{\tablecaptionfont%
Table~\the\tableno. #1}
\vskip-0.4 true cm}

\def\thintablerule{\hrule height0.4pt}

\tenpoint

\immediate\openin\bibitemlist=biblist
\ifeof\bibitemlist\immediate\closein\bibitemlist
\else\immediate\closein\bibitemlist
\input biblist \fi


\def\thismonth{\ifcase\month\or
January\or February\or March\or April\or May\or June\or
July\or August\or September\or October\or November\or December\fi}

\input epsf
\epsfclipon



\def\rmd{{\rm d}}

\def\rme{{\rm e}}
\def\rmO{{\rm O}}


\def\rz{{\Bbb R}}


\def\proof{\noindent{\sl Proof:}\kern0.6em}

\def\frac#1#2{\hbox{$#1\over#2$}}
\def\dual{\mathstrut^*\kern-0.1em}

\def\lvec#1{\setbox0=\hbox{$#1$}
    \setbox1=\hbox{$\scriptstyle\leftarrow$}
    #1\kern-\wd0\smash{
    \raise\ht0\hbox{$\raise1pt\hbox{$\scriptstyle\leftarrow$}$}}
    \kern-\wd1\kern\wd0}
\def\rvec#1{\setbox0=\hbox{$#1$}
    \setbox1=\hbox{$\scriptstyle\rightarrow$}
    #1\kern-\wd0\smash{
    \raise\ht0\hbox{$\raise1pt\hbox{$\scriptstyle\rightarrow$}$}}
    \kern-\wd1\kern\wd0}
\def\slash#1{\setbox0=\hbox{$#1$}\setbox1=\hbox{$\kern1pt/$}
    #1\kern-\wd0\kern1pt/\kern-\wd1\kern\wd0}


\def\nabstar#1{{\nabla\kern0.5pt\smash{\raise 4.5pt\hbox{$\ast$}}
               \kern-5.5pt_{#1}}}

\def\drvstar#1{{\partial\kern0.5pt\smash{\raise 4.5pt\hbox{$\ast$}}
               \kern-6.0pt_{#1}}}

\def\ldrvstar#1{{\lvec{\,\partial}\kern-0.5pt\smash{\raise 4.5pt\hbox{$\ast$}}
               \kern-5.0pt_{#1}}}


\def\MSbar{\overline{\rm MS\kern-0.5pt}\kern0.5pt}



\def\psibar{\overline{\psi}{\vphantom{\psi}}\kern-0.6pt}
\def\chibar{\overline{\chi}{\vphantom{\chi}}\kern-0.6pt}
\def\Vbar{\kern1.0pt\overline{\kern-1.0ptV\kern-1.0pt}\kern1.0pt{\vphantom{V}}}
\def\cbar{\bar{c}}
\def\ctilde{\tilde{c}{\kern1pt\vphantom{c}}}
\def\cbartilde{\tilde{\cbar}{\kern1pt\vphantom{c}}}
\def\dbar{\bar{d}}
\def\dtilde{\tilde{d}{\kern1pt\vphantom{d}}}
\def\dbartilde{\tilde{\dbar}{\kern1pt\vphantom{d}}}
\def\ren#1{#1_{\hbox{\sixrm R}}}

\def\Ar{\ren{A}}


\def\diracstar#1#2{
    \setbox0=\hbox{$\gamma$}\setbox1=\hbox{$\gamma_{#1}$}
    \gamma_{#1}\kern-\wd1\kern\wd0
    \smash{\raise4.5pt\hbox{$\scriptstyle#2$}}}


\def\SUn{{\rm SU}(N)}

\def\sun{\frak{su}(N)}
\def\tr{{\rm tr}}


\def\Sgf{S_{\rm gf}}
\def\Sgh{S_{c\cbar}}
\def\Sfl{S_{\rm fl}}
\def\Sflgh{S_{d\dbar}}
\def\Stot{S_{\rm tot}}
\def\Sbc{S_{\rm bc}}


\def\vbulk#1{X^{(#1)}}

\def\Dtilde{\tilde{D}}


\def\dbrs{\delta}


\def\eps{\epsilon}

\def\msbar{{\rm\overline{MS\kern-0.05em}\kern0.05em}}
\def\gbar{\bar{g}}

%
\rightline{CERN-PH-TH/2010-290}
\rightline{MPP-2010-165}
\vskip1.2cm 
\centerline{\Bigrm
Perturbative analysis of the gradient flow 
}
\vskip0.3cm
\centerline{\Bigrm
in non-abelian gauge theories}
\vskip 0.6 true cm
\centerline{\bigrm Martin L\"uscher}
\vskip1.5ex
\centerline{{\it CERN, Physics Department, 1211 Geneva 23, Switzerland}}
\vskip 0.4 true cm
\centerline{\bigrm Peter Weisz}
\vskip1ex
\centerline{\it Max-Planck-Institut f\"ur Physik, 80805 Munich, Germany}
\vskip 0.8 true cm
\thintablerule
\vskip 2.0ex
\ninepoint
\leftline{\bf Abstract}
\vskip 1.0ex\noindent
The gradient flow in non-abelian gauge theories 
on $\rz^4$
is defined by a local diffusion equation that evolves
the gauge field as a function of the flow time
in a gauge-covariant manner.
Similarly to the case of the Langevin equation,
the correlation functions of the time-dependent field 
can be expanded in perturbation theory, 
the Feynman rules being those of a 
renormalizable field theory on $\rz^4\times[0,\infty)$.
For any matter multiplet and to all loop orders,
we show that the correlation functions are finite, i.e.~do not require 
additional renormalization,
once the theory in four dimensions 
is renormalized in the usual way. 
The flow thus maps the gauge field to 
a one-parameter family of smooth renormalized fields.
\vskip 2.0ex
\thintablerule

\tenpoint

\vskip-0.3cm

\section 1. Introduction

The physics described by non-abelian gauge theories
can be studied in many ways. Depending on the 
context and the questions to be answered, 
the desired information is extracted from 
correlation functions of local fields, 
expectation values of Wilson loops or the Schr\"o\-din\-ger 
functional, for example. 

As explained in refs.~[\ref{WilsonFlow},\ref{Villasimius}],
the gradient flow provides further opportunities 
to probe these theories
and allows some of their otherwise elusive properties
to be understood
(in lattice gauge theory, the gradient flow is referred to
as the Wilson flow, 
while in the mathematical literature it 
is commonly known as the Yang--Mills gradient flow).
Evidently, physically meaningful probes must be safe of ultra-violet
divergences or must be such that these can be canceled by
a well-defined renormalization procedure. 

In the case of the gradient flow,
there is some evidence that the gauge field generated by the
flow does not require renormalization [\ref{WilsonFlow},\ref{Villasimius}],
but a formal proof of the absence of ultra-violet divergences at 
positive flow time has not been given so far.
The aim of the present paper is to fill this gap
through an all-order analysis of the flow in perturbation theory.
With respect to the closely related case of the renormalization of
the Langevin equation discussed by 
Zinn--Justin and Zwanziger [\ref{Zinn},\ref{ZinnZwanziger}],
there are two important differences, one being the absence of 
the noise term in the flow equation and the other the fact that
the initial distribution of the gauge field 
(which is given by the functional integral of the theory considered) 
is not ignored.

The perturbative analysis presented in this paper applies to 
renormalizable gauge theories with any compact simple gauge group 
and any matter multiplet.
However, in order to simplify the discussion as much as
possible, only the pure $\SUn$ gauge theory with dimensional
regularization will be considered, the generalization
to other cases being straightforward (see sect.~8).

In the following three sections, the Feynman rules for the correlation 
functions of the gauge
field generated by the gradient flow are derived and are
shown to be those of a local field theory with an extra dimension
(the flow time). 
The finiteness of the correlation functions
at positive flow time can then be established 
using power-counting and the BRS symmetry (sects.~6,7),
but for illustration the divergent parts of a set of 
one-loop diagrams are first worked out
in sect.~5.

\section 2. Iterative solution of the flow equation

In this section, we introduce the gradient flow
and derive the Feynman rules for the associated flow-line diagrams.
The pure $\SUn$ gauge theory in $D=4-2\eps$ 
euclidean dimensions is considered and 
the fundamental gauge field $A_{\mu}(x)$ is normalized so that 
its action at bare coupling $g_0$ is given by
\equation{
  S=-{1\over2g_0^2}\int\rmd^Dx\,\tr\{F_{\mu\nu}(x)F_{\mu\nu}(x)\},
  \enum
  \nexteq{2.0ex}
  F_{\mu\nu}=\partial_{\mu}A_{\nu}-\partial_{\nu}A_{\mu}+[A_{\mu},A_{\nu}]
  \enum
}
(see appendix A for unexplained notation).

\subsection 2.1 Definition of the gradient flow

The gradient flow evolves the gauge field as a function 
of a parameter $t\geq0$ that is referred to as the flow time.
Starting from the fundamental gauge field,
\equation{
  \left.B_{\mu}\right|_{t=0}=A_{\mu},
  \enum
}
the time-dependent field $B_{\mu}(t,x)$
is determined by the differential equation
\equation{
  \partial_tB_{\mu}=D_{\nu}G_{\nu\mu}+\alpha_0D_{\mu}\partial_{\nu}B_{\nu},
  \enum
  \nexteq{2ex}
  G_{\mu\nu}=\partial_{\mu}B_{\nu}-\partial_{\nu}B_{\mu}+[B_{\mu},B_{\nu}],
  \qquad
  D_{\mu}=\partial_{\mu}+[B_{\mu},\,\cdot\;].
  \enum
}
The name ``gradient flow'' derives from the fact that first term on 
the right of eq.~(2.4) is proportional to the gradient of the gauge
action along the flow. 
Note that neither the initial condition (2.3) nor the 
flow equation (2.4) involve the gauge coupling.

The second term on the right of eq.~(2.4) is included in order to 
damp the evolution of the gauge degrees of freedom of the 
field. As in the case of the Langevin equation [\ref{ZinnZwanziger}], 
some technicalities in the perturbative analysis of the flow 
can be avoided in this way without affecting the evolution
of the gauge-invariant observables.
The latter are in fact independent
of the parameter $\alpha_0$, since
the solutions of eq.~(2.4) obtained at different values
of $\alpha_0$ are related by 
a (time-dependent) gauge transformation [\ref{WilsonFlow}].

\subsection 2.2 Expansion in powers of the fundamental gauge field

Equation (2.4) may be split into a linear and remainder part
according to
\equation{
  \partial_tB_{\mu}=\partial_{\nu}\partial_{\nu}B_{\mu}
  +(\alpha_0-1)\partial_{\mu}\partial_{\nu}B_{\nu}+R_{\mu},
  \enum
  \nexteq{2ex}
  R_{\mu}=2[B_{\nu},\partial_{\nu}B_{\mu}]-
  [B_{\nu},\partial_{\mu}B_{\nu}]+(\alpha_0-1)[B_{\mu},\partial_{\nu}B_{\nu}]
  +[B_{\nu},[B_{\nu},B_{\mu}]].
  \enum
}
The linearized equation can be solved using the heat kernel
\equation{
  K_t(z)_{\mu\nu}=
  \int_p{\rme^{ipz}\over p^2}\bigl\{
  (\delta_{\mu\nu}p^2-p_{\mu}p_{\nu})\rme^{-tp^2}+
  p_{\mu}p_{\nu}\rme^{-\alpha_0tp^2}\bigr\},
  \enum
}
where
\equation{
  \int_p=\int{\rmd^Dp\over(2\pi)^D}.
  \enum
}
Taking the boundary condition (2.3) into account, the flow equation
may then be cast in the integral form
\equation{
  B_{\mu}(t,x)=\int\rmd^Dy\,\Bigl\{
  K_t(x-y)_{\mu\nu}A_{\nu}(y)
  +\int_0^t\rmd s\,K_{t-s}(x-y)_{\mu\nu}R_{\nu}(s,y)\Bigr\}.
  \enum
}
From this representation the retarded character of the equation
is evident and it is also quite
clear that the sensitivity to the initial value of the field dies away as
$t$ increases, although only slowly so at small momenta
($\alpha_0$ is assumed to be positive).

When passing to momentum space,
\equation{
  B_{\mu}(t,x)=\int_p\rme^{ipx}\tilde{B}_{\mu}(t,p),
  \enum
}
the integral equation (2.10) becomes
\equation{
  \tilde{B}_{\mu}(t,p)=\tilde{K}_t(p)_{\mu\nu}\tilde{A}_{\nu}(p)+
  \int_0^t\rmd s\,\tilde{K}_{t-s}(p)_{\mu\nu}\tilde{R}_{\nu}(s,p).
  \enum
}
It is helpful at this point to introduce
the vertices $\vbulk{2,0}$ and $\vbulk{3,0}$ through
\equation{
  \tilde{R}_{\mu}^a(t,p)=\sum_{n=2}^3
  {1\over n!}\int_{q_1}\ldots\int_{q_n}(2\pi)^D\delta(p+q_1+\ldots+q_n)
  \noenum
  \nexteq{2.5ex}
  {\phantom{\tilde{R}_{\mu}(t,p)=}}
  \times\vbulk{n,0}(p,q_1,\ldots,q_n)^{ab_1\ldots b_n}_{\mu\nu_1\ldots\nu_n}
  \tilde{B}^{b_1}_{\nu_1}(t,-q_1)\ldots\tilde{B}^{b_n}_{\nu_n}(t,-q_n)
  \enum
}
and the requirement that they are totally symmetric in the
momentum-index combinations $(q_1,\nu_1,a_1),\ldots,(q_n,\nu_n,a_n)$.
The solution of the integral equation in powers of the 
fundamental gauge field,
\equation{
  \tilde{B}_{\mu}^a(t,p)=\tilde{K}_t(p)_{\mu\nu}\tilde{A}_{\nu}^a(p)
  +\frac{1}{2}\int_0^t\rmd s\,\tilde{K}_{t-s}(p)_{\mu\nu}
  \int_{q,r}(2\pi)^D\delta(p-q-r)
  \noenum
  \nexteq{2.5ex}
  {\phantom{\tilde{B}_{\mu}^a(t,p)=}}
  \times
  \vbulk{2,0}(p,-q,-r)^{abc}_{\nu\rho\sigma}
  \tilde{K}_{s}(q)_{\rho\delta}\tilde{K}_{s}(r)_{\sigma\tau}
  \tilde{A}^b_{\delta}(q)\tilde{A}^c_{\tau}(r)+\ldots,
  \enum
}
is then obtained through iteration, i.e.~by 
recursively inserting the equation on the right of itself.

The vertices $\vbulk{2,0}$ and $\vbulk{3,0}$ are given explicitly
in appendix B. Note that the momentum-index combination 
$(p,\mu,a)$ plays a special r\^ole in eq.~(2.13). In particular,
the vertices are symmetric only in their other arguments.

\topinsert
\vbox{
\vskip0.0cm
\centerline{\epsfysize=2.7cm\epsfbox{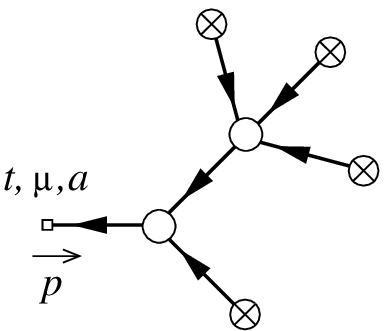}}
\vskip0.5cm
\figurecaption{%
The diagrams contributing to $\tilde{B}^a_{\mu}(t,p)$
are directed tree graphs with
a single external line (a little square is drawn at the end of this line).
Flow lines always start from a one-point vertex
(circle with a cross)
or a flow vertex (circle) and end at another flow vertex 
if the line is not the external one. Each flow vertex has 
one outgoing flow line that corresponds to the momentum-index
combination $(p,\mu,a)$ in eq.~(2.13).
}
}
\endinsert

\subsection 2.3 Flow-line diagrams

The terms contributing to the expansion (2.14) in powers of the 
fundamental gauge field can be graphically represented by
Feynman diagrams (see fig.~1).
There are two kinds of vertices in these diagrams,
the flow vertices $\vbulk{2,0}$ and $\vbulk{3,0}$ introduced
in the previous subsection and the one-point vertex
\equation{
  \raise-0.54cm\hbox{\epsfxsize=1.0cm\epsfbox{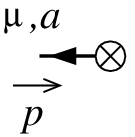}}%
  \;=\;\tilde{A}^a_{\mu}(p).
  \enum
}
Each flow vertex is inserted at some flow time that 
is integrated from zero to infinity, while the one-point
vertices reside at time zero.
The vertices are connected through the flow lines
\equation{
  \raise-0.68cm\hbox{\epsfxsize=2.5cm\epsfbox{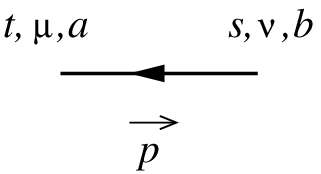}}%
  =\;\delta^{ab}\theta(t-s)\tilde{K}_{t-s}(p)_{\mu\nu},
  \enum
}
where $t$ and $s$ are the flow times at the endpoints of the line.
In view of the retarded nature
of the propagator (2.16), the flow time increases
from zero at the one-point vertices to time $t$
at the outer end of the external line as one follows
the arrows in the diagram.
In particular, 
the times associated with the vertices 
are effectively integrated only up to $t$ (rather than infinity).

As far the momenta, the index contractions and the symmetry factors 
are concerned, the Feynman rules are the usual ones. 
It is then not difficult to show that the sum of all
diagrams solves the flow equation (2.12) order
by order in the fundamental gauge field.

\section 3. Perturbation theory

The $n$-point 
correlation functions of the field $B^a_{\mu}(t,x)$ 
can be computed in perturbation theory by expanding the 
field in powers of the fundamental field, as explained sect.~2, 
and by calculating the correlation functions of the latter 
in the $\SUn$ gauge theory as usual.
We now show that the correlation functions can be 
directly obtained from 
a set of Feynman rules in $D+1$ dimensions.

\subsection 3.1 Gauge fixing

The flow equation (2.4) 
is invariant under the infinitesimal transformation
\equation{
  \delta B_{\mu}=D_{\mu}\omega,
  \enum
}
provided the time-dependence of $\omega(t,x)\in\sun$ 
is such that
\equation{
  \partial_t\omega=\alpha_0D_{\mu}\partial_{\mu}\omega.
  \enum
}
Since the initial value of $\omega$ is unconstrained, these transformations
generate the full gauge group at flow time $t=0$ and thus extend the 
gauge symmetry of the $\SUn$ gauge theory to all flow times.

The symmetry can be fixed as usual by including the
gauge-fixing term
\equation{
  \Sgf=-{\lambda_0\over g_0^2}\int\rmd^Dx\,
  \tr\{\partial_{\mu}A_{\mu}(x)\partial_{\nu}A_{\nu}(x)\}
  \enum  
}
and the associated ghost action
\equation{
  \Sgh=-{2\over g_0^2}\int\rmd^Dx\,
  \tr\{\partial_{\mu}\cbar(x)D_{\mu}c(x)\}
  \enum
}
in the total action of the theory, where $c$ and $\cbar$ are the 
Faddeev--Popov ghost fields.
As far as the $n$-point 
correlation functions of the fundamental gauge field
and the ghosts are concerned, the Feynman rules are then the standard
ones. Evidently, since the transformation (3.1) is
an infinitesimal gauge variation, the expectation values of 
gauge-invariant expressions in the field generated by the flow
are independent of the gauge parameter $\lambda_0$.

\subsection 3.2 Gauge-field propagator

The way in which the flow-line diagrams combine with the 
Feynman diagrams of the underlying theory is best
explained by considering the two-point function
of the time-dependent gauge field.
To leading order in the gauge coupling,
the flow-line diagram 
\equation{
  \raise-0.55cm\hbox{\epsfxsize=3.5cm\epsfbox{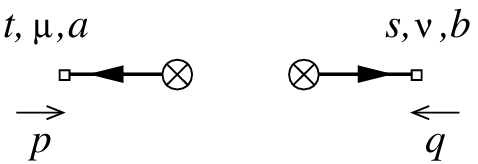}}
  \enum
}
is the only one that contributes to the correlation function.
The contraction of the gauge fields at the one-point 
vertices then shows that
\equation{
  \langle \tilde{B}^a_{\mu}(t,p)\tilde{B}^b_{\nu}(s,q)\rangle=
  (2\pi)^D\delta(p+q)\delta^{ab}g_0^2\Dtilde_{t+s}(p)_{\mu\nu}+\rmO(g_0^4),
  \enum
  \nexteq{2.0ex}
  \Dtilde_t(p)_{\mu\nu}=
  {1\over(p^2)^2}
  \bigl\{
  (\delta_{\mu\nu}p^2-p_{\mu}p_{\nu})\rme^{-tp^2}+
  \lambda_0^{-1}p_{\mu}p_{\nu}\rme^{-\alpha_0tp^2}\bigr\}.
  \enum
}
This formula includes the mixed propagator
\equation{
  \langle \tilde{A}^a_{\mu}(p)\tilde{B}^b_{\nu}(s,q)\rangle=
  \langle \tilde{B}^a_{\mu}(0,p)\tilde{B}^b_{\nu}(s,q)\rangle
  \enum
}
as well as the two-point function of the gauge field
at flow time zero.

Since all three propagators are given by the same analytic expression,
the same graphical symbol
\equation{
  \raise-0.65cm\hbox{\epsfxsize=2.5cm\epsfbox{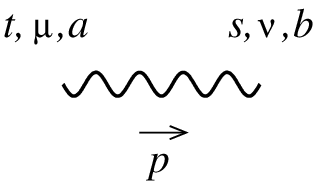}}%
  =\;\delta^{ab}g_0^2\Dtilde_{t+s}(p)_{\mu\nu}
  \enum
}
may be used for them. Note that the contraction of the one-point vertices
always has the effect of converting the terminal flow lines
to gauge-field lines.
If one starts from the flow-line diagram
\equation{
  \raise-0.70cm\hbox{\epsfxsize=4.5cm\epsfbox{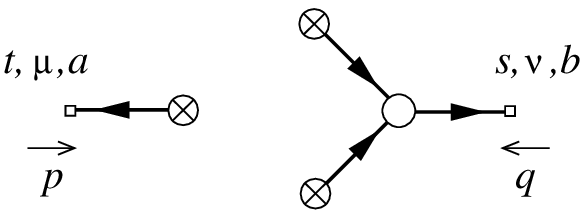}}
  \enum
}
instead of the diagram (3.5), for example, the 
leading-order contribution is obtained by substituting 
the tree-level diagram for the correlation function
of the gauge fields at the three one-point vertices.
This leads to a diagram
\equation{
  \raise-0.61cm\hbox{\epsfxsize=3.6cm\epsfbox{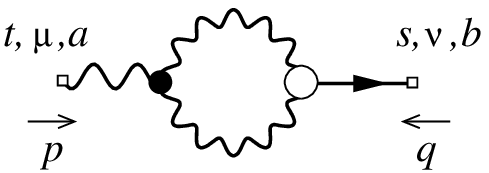}}
  \enum
}
that has an ordinary three-point vertex (filled circle) 
and a flow vertex. Two of~the lines 
attached to the latter are gauge-field lines instead 
of flow lines, but the expression for the vertex is the same as before.

\subsection 3.3 Feynman rules in $D+1$ dimensions

The flow time will now be interpreted as an additional 
space-time coordinate. Since only non-negative times are considered,
the $D+1$ dimensional space is a half-space with
a $D$ dimensional boundary at flow time zero.
The $\SUn$ gauge theory 
lives at the boundary, while the field generated by the 
gradient flow extends to the extra dimension.

From this point of view, the ordinary and the flow vertices represent 
boundary and bulk interaction terms, respectively, while
the propagation of the fields in $D+1$ dimensions is described
by the gauge-field propagator (3.9) and the flow propagator (2.16).
The ghost-field propagator
\equation{
  \langle\ctilde^a(p)\cbartilde^b(q)\rangle=
  (2\pi)^D\delta(p+q)\delta^{ab}g_0^2\Dtilde(p)+\rmO(g_0^4),
  \qquad  
  \Dtilde(p)={1\over p^2},
  \enum
}
on the other hand, is defined at flow time zero only.
Graphically it is represented through a dotted line,
\equation{
  \raise-0.65cm\hbox{\epsfxsize=1.8cm\epsfbox{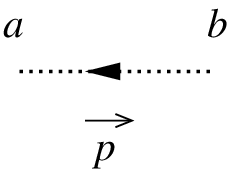}}%
  =\;\delta^{ab}g_0^2\Dtilde(p),
  \enum
}
where the arrow is drawn in the direction 
from $\cbar$ to $c$.

\topinsert
\vbox{
\vskip0.0cm
\centerline{\epsfxsize=10.0cm\epsfbox{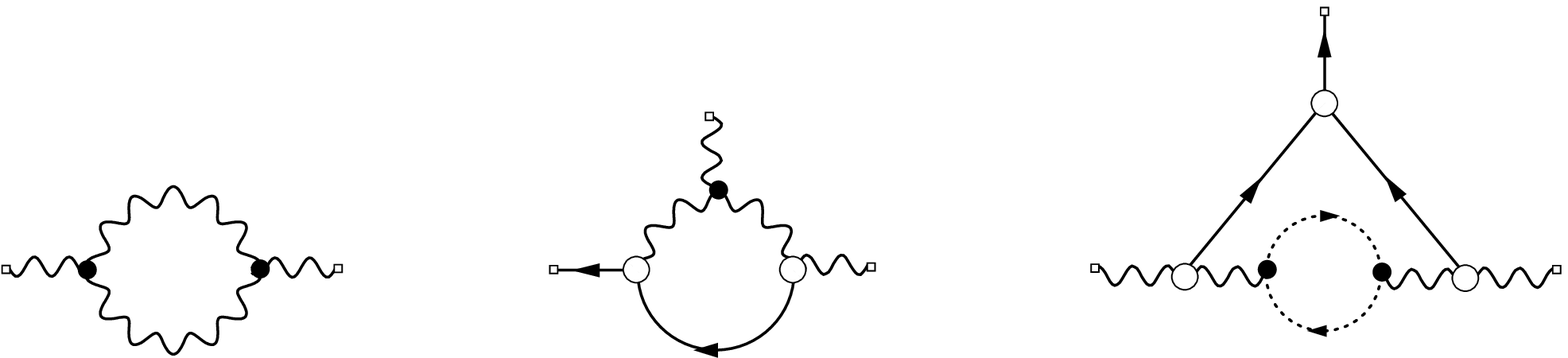}}
\vskip0.5cm
\figurecaption{%
Examples of diagrams contributing to the two- and three-point 
functions of the time-dependent gauge field. 
All diagrams are built from flow propagators
(directed solid lines), gauge-field propagators (wiggly lines),
ghost-field propagators (directed dotted lines), flow vertices (open circles)
and ordinary vertices at flow time zero (filled circles). The little
square at the end of the external lines indicates that
they are not amputated. 
}
}
\endinsert

It should be quite clear at this point that the $n$-point functions
of the field generated by the gradient flow are given by 
Feynman diagrams in $D+1$ dimensions with the graphical
elements listed in the caption of fig.~2. 
Some special features of the Feynman rules are worth pointing out:

\vskip1.0ex
\noindent
(a) Each flow vertex has exactly one outward-directed flow line corresponding
to the first momentum-index combination in eq.~(2.13). The other lines
attached to these vertices may be gauge-field lines or ingoing flow lines.

\vskip1.0ex
\noindent
(b) Flow lines must start at a 
flow vertex and are either external or end at another flow vertex. 
The gauge-field lines, on the other hand, can start and end at
both the flow vertices and the ordinary vertices.

\vskip1.0ex
\noindent
(c) The flow vertices are inserted at some flow time which is integrated
from zero to infinity. All other vertices are at flow time zero. 
The flow times on which the propagators depend are the ones at
the endpoints of the corresponding lines.

\vskip1.0ex
\noindent
(d) Diagrams with closed flow-line loops are set to zero. 
This rule derives from the fact that flow-line diagrams are tree 
diagrams and that the contraction of the gauge fields at the one-point 
vertices never leads to new flow lines.

\section 4. Field theory ${\mib D}{\bf+1}$ dimensions

The Feynman rules for the correlation functions of the time-dependent
gauge field obtained in the previous section
are those of a local field theory in $D+1$ dimensions [\ref{ZinnZwanziger}].
It is possible to show this through somewhat formal
functional-integral mani\-pulations, but one can also adopt
a purely algebraic point of view, where the action in $D+1$ dimensions 
merely serves as a generating function for the Feynman rules.

\subsection 4.1 Action

Except for the boundary condition (2.3),
the field $B_{\mu}(t,x)$ will now be considered to be an 
independent field.
One also needs to introduce a Lagrange-multiplier 
field $L_{\mu}(t,x)=L_{\mu}^a(t,x)T^a$
with purely imaginary components. 
The action of the theory in $D+1$ dimensions is then given by
[\ref{ZinnZwanziger}]
\equation{
   \Stot=S+\Sgf+\Sgh+\Sfl,
   \enum
   \nexteq{2.5ex}
   \Sfl=-2\int_0^{\infty}\rmd t\int\rmd^Dx\, 
   \tr\bigl\{L_{\mu}(t,x)
   \bigl(\partial_tB_{\mu}-
         D_{\nu}G_{\nu\mu}-\alpha_0D_{\mu}\partial_{\nu}B_{\nu}
   \bigr)(t,x)\bigr\}.
   \enum
}
In this framework, the flow equation (2.4) coincides with 
the field equation obtained by varying the action with respect to the
Lagrange-multiplier field. Note that the latter is not required to 
satisfy any particular boundary conditions.

\subsection 4.2 Propagators

The quadratic part of the action is the sum of 
\equation{
   \int_0^{\infty}\rmd t\int\rmd^Dx\, L^a_{\mu}(t,x)
   \bigl(
   \partial_tB^a_{\mu}-\partial_{\nu}\partial_{\nu}B^a_{\mu}
   -(\alpha_0-1)\partial_{\mu}\partial_{\nu}B^a_{\nu}
   \bigr)(t,x)
   \enum
}
and the quadratic part of the action in $D$ dimensions. 
The bulk fields $B_{\mu}$ and $L_{\mu}$
thus couple to each other only, but there is also
an implicit coupling to the fundamental gauge field
through the boundary condition (2.3). 
This complication can easily be overcome by substituting
\equation{
   B_{\mu}(t,x)=\int\rmd^Dy\,K_t(x-y)_{\mu\nu}A_{\nu}(y)+b_{\mu}(t,x).
   \enum
}
The field $b_{\mu}$ then satisfies homogenous boundary conditions,
while the first term in eq.~(4.4) solves the linearized flow equation
and thus drops out in the action (4.3).

It follows from these remarks that the propagators of 
the fundamental gauge field and the ghost fields
coincide with the expressions given in the previous section. 
All other two-point functions vanish except for
\equation{
   \left.\bigl\langle 
   b^a_{\mu}(t,x)L^b_{\nu}(s,y)
   \bigr\rangle\right|_{\rm leading\;order}
   =\delta^{ab}H(t,x;s,y)_{\mu\nu},
   \enum
}
which is determined by the field equation
\equation{
   \bigl\{\delta_{\mu\rho}\partial_t-
   \delta_{\mu\rho}\partial_{\sigma}\partial_{\sigma}
   -(\alpha_0-1)\partial_{\mu}\partial_{\rho}\bigr\}
   H(t,x;s,y)_{\rho\nu}=\delta_{\mu\nu}\delta(t-s)\delta(x-y)
   \enum
}
and the boundary condition $\left.H_{\mu\nu}(t,x;s,y)\right|_{t=0,s>0}=0$.
The unique solution of these equations is
\equation{
   H(t,x;s,y)_{\mu\nu}=\theta(t-s)K_{t-s}(x-y)_{\mu\nu}
   \enum
}
and the $bL$ propagator is thus seen to coincide with the flow 
propagator. 

The two-point functions involving the $B$ field
may finally be calculated by combining eq.~(4.4) with the results
obtained so far. Since the $AL$ and the $bb$ propagators vanish, one 
quickly finds that the $BL$ propagator is equal to the flow propagator
and that the $BB$ and the $AB$ propagators are
equal to the expressions given in the previous section.

\subsection 4.3 Vertices

The theory described by the action (4.1) has the vertices of the 
theory in $D$ dimensions plus those deriving from the bulk action
$\Sfl$. Recalling eqs.~(2.6) and (2.13), it is straightforward to show that
the $LB^2$ and $LB^3$ vertices generated by 
the interaction part of the latter,
\equation{
   \left.\Sfl\right|_{\rm interaction}=
   2\int_0^{\infty}\rmd t\int\rmd^Dx\, 
   \tr\bigl\{L_{\mu}(t,x)R_{\mu}(t,x)\bigr\},
   \enum
}
coincide with the vertices $\vbulk{2,0}$ and $\vbulk{3,0}$. 

All vertices and propagators of the 
Feynman rules of the previous sections are thus recovered.
A somewhat unusual feature of these rules is the presence
of a field (the Lagrange multiplier) that propagates
only through its mixing with the other fields.
The algebraic structure of the Feynman rules
is however entirely standard and does not involve any 
prescriptions apart from the ones deriving from the expansion of 
the action in powers of the fields.

\topinsert
\vbox{
\vskip0.0cm
\centerline{\epsfxsize=6.5cm\epsfbox{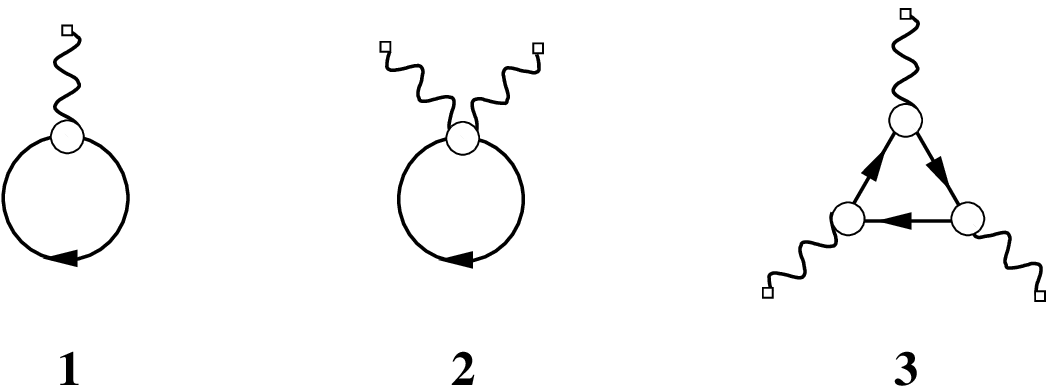}}
\vskip0.3cm
\figurecaption{%
Examples of diagrams with flow-line loops. All such diagrams 
vanish as a consequence of the retarded nature of 
the flow propagator and the rules of dimensional regularization
[\ref{ZinnZwanziger}].
}
}
\endinsert

\subsection 4.4 Flow lines and flow-line loops

Flow lines represent the $BL$ propagator. They can start
and end at the flow vertices but are never attached
to an ordinary vertex. Outward- and inward-directed 
external flow lines are external $B$ and $L$ lines,
respectively. 

Diagrams with closed flow lines comply with these rules,
but should be absent if
a complete matching with the Feynman rules of 
the previous section is to be achieved
(cf.~point (d) at the end of sect.~3).
Such diagrams are in fact equal to zero [\ref{ZinnZwanziger}].
The diagrams 1 and 2 in fig.~3, for example, vanish because
dimensional regularization sets the momentum integral to zero.
If there are two or more vertices in the 
loop, as in diagram 3, the time integrations
vanish, because the flow propagator is retarded and thus forces
the flow times at the vertices to be squeezed to a range of 
measure zero (singularities in the time 
coordinates are excluded in view of
the regularization of the momentum integral).

If a lattice regularization is used, the vanishing of diagram 2 may
not be guaranteed, but one can always include a ghost
field in the action that cancels the flow-line loops algebraically,
i.e.~at the level of the Feynman integrands [\ref{Zinn}]. 
With dimensional regularization,
however, this device is not needed and is therefore omitted here.

\section 5. Sample calculation at one-loop order

For illustration, the divergent parts of the one-loop diagrams
that contribute to the gauge-field two-point functions are 
worked out in this section. The parameter and field 
renormalization required at flow time zero
is then seen to cancel the singularities at all flow times.

\topinsert
\vbox{
\vskip0.0cm
\centerline{\epsfxsize=9.6cm\epsfbox{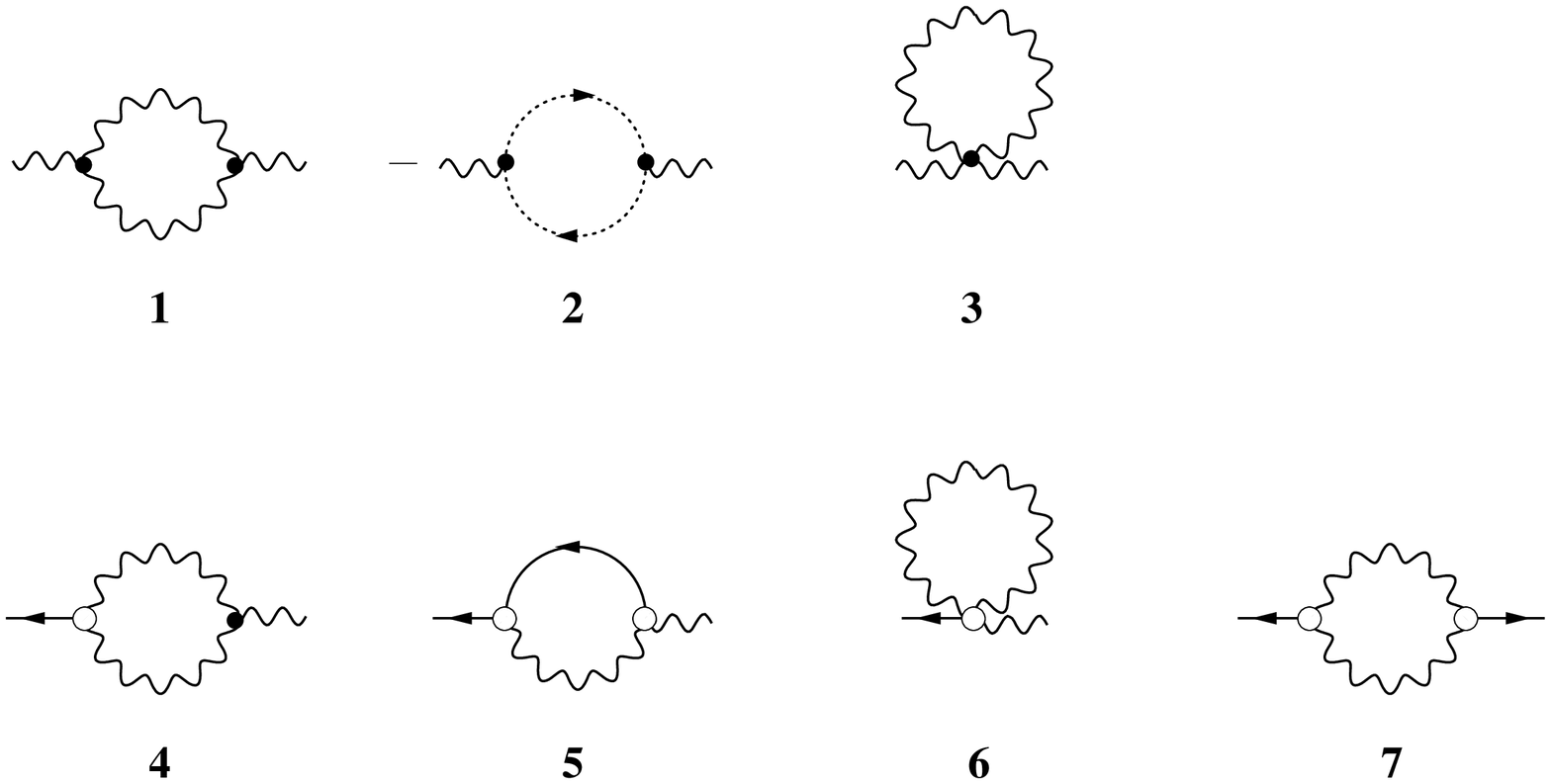}}
\vskip0.5cm
\figurecaption{%
One-loop diagrams contributing to the 
$AA$ ($1-3$), $LA$ ($4$), $LB$ ($5,6$) and 
$LL$ ($7$) vertex functions.
The time-momentum-index combinations at the
external legs on the left and right 
are $(t,p,\mu,a)$ and $(s,q,\nu,b)$, respectively
(the flow times are absent in the case of the $A$ legs). 
}
}
\endinsert

\subsection 5.1 Computation of self-energy diagrams

There are only few self-energy diagrams at one-loop order (see fig.~4).
The diagrams $1-3$ coincide with the ones contributing to the 
two-point function of the gauge field at flow time zero. 
We therefore merely quote the known result
\equation{
   \left.\Gamma^{(1)}_{AA}(p)_{\mu\nu}^{ab}
   \right|_{\rm pole}
   =\delta^{ab}(\delta_{\mu\nu}p^2-p_{\mu}p_{\nu})
   {N\over16\pi^2\eps}\left({13\over6}-{1\over 2\lambda_0}\right)
   \enum
}
for the sum of their singular parts.

In the case of diagram $4$,
the Feynman integrand is a linear combination
of terms of the form
\equation{
   \delta^{ab}
   {Q(k,p)_{\mu\nu}\over \{k^2(k+p)^2\}^2}
   \rme^{-t\{uk^2+v(k+p)^2\}},
   \qquad u,v\in\{1,\alpha_0\},
   \enum
}
where $k$ is the loop momentum and $Q(k,p)_{\mu\nu}$ 
a homogeneous polynomial in $k$ and $p$ of degree $6$
(the momentum-conservation factor $(2\pi)^D\delta(p+q)$ 
is now always suppressed). This diagram is finite
at all $t>0$ and in any dimension $D$,
but eventually it will be multiplied by
the external propagators and must then be integrated
over $t$ from $0$ to infinity. The behaviour
of the integral small $t$ thus matters and can give rise 
to singularities at $D=4$. 

If $f(t)$ is any smooth test function (an external propagator, for 
example), the time integral
\equation{
   I(k,p)=\int_{0}^{\infty}\rmd t\, f(t)\rme^{-t\{uk^2+v(k+p)^2\}}
\enum
}
can be worked out in an asymptotic series at large $k$,
the leading term being 
\equation{
   I(k,p)={f(0)\over(u+v)k^2}+\rmO(k^{-3}).
\enum
}
The integration over the flow time $t$ thus improves the degree of
divergence of the momentum integral. In particular, the first term
in the decomposition
\equation{
   I(k,p)=\left\{I(k,p)-{f(0)\over(u+v)(k+r)^2}\right\}+
   {f(0)\over(u+v)(k+r)^2}
   \enum
}
(where $r$ is an arbitrary external momentum) makes the integral
convergent at $D=4$.
One is then left with a logarithmic divergence and 
therefore a pole singularity equal to
the one of the ordinary Feynman integral
\equation{
   \delta^{ab}\delta(t){1\over u+v}\int_k
   {Q(k,0)_{\mu\nu}\over(k^2)^4(k+r)^2},
   \enum
}
which is easily evaluated using standard techniques.

Proceeding in this way, one obtains
\equation{
   \left.\Gamma^{(1)}_{LA}(t,p)_{\mu\nu}^{ab}
   \right|_{\rm pole}
   =g_0^2\delta^{ab}\delta(t)\delta_{\mu\nu}
   {N\over16\pi^2\eps}
   \left({3\over4}+{3\over 4\lambda_0}\right),
   \enum
   \nexteq{3.5ex}
   \left.\Gamma^{(1)}_{LB}(t,s,p)_{\mu\nu}^{ab}
   \right|_{\rm pole}
   =g_0^2\delta^{ab}\delta(t)\delta(s)\delta_{\mu\nu}
   {N\over16\pi^2\eps}\left(-{1\over 2\lambda_0}\right),
   \enum
}
for the divergent parts of the $LA$ and $LB$ vertex functions. 
The diagram $7$
and thus the $LL$ vertex function are finite at $D=4$.

\subsection 5.2 Renormalization

The bare coupling and gauge-fixing parameter are 
related to the renormalized parameters $g$ and $\lambda$ 
through
\equation{
   g_0^2=\mu^{2\eps}g^2Z,
   \qquad 
   \lambda_0=\lambda Z_3^{-1},
   \enum
}
where $\mu$ is the normalization mass.
To one-loop order, the renormalization constants 
$Z$ and $Z_3$ are given by
\equation{
   Z=1-{b_0\over\eps}g^2+\rmO(g^4),
   \qquad 
   b_0={N\over16\pi^2}{11\over3},
   \enum
   \nexteq{3.0ex}
   Z_3=1+{c_0\over\eps}g^2+\rmO(g^4),
   \qquad 
   c_0={N\over16\pi^2}\left({13\over6}-{1\over2\lambda}\right),
   \enum
}
up to scheme-dependent finite terms.

The renormalization of the fundamental gauge field, 
\equation{
   A_{\mu}^a=Z^{1/2}Z_3^{1/2}(\Ar)^a_{\mu},
   \enum
}
involves both renormalization constants
as a result of the unconventional normalization conventions 
adopted in this paper. For reasons explained in sect.~7, 
the ghost fields are renormalized asymmetrically according to 
\equation{
   c^a=\tilde{Z}_3Z^{1/2}Z_3^{1/2}(\ren{c})^a,
   \qquad
   \cbar^a=Z^{1/2}Z_3^{-1/2}(\ren{\cbar})^a,
   \enum
} 
where $\tilde{Z}_3$ is the usual ghost renormalization constant.
Note that eq.~(5.13) is equivalent
to the standard renormalization prescription for the 
correlation functions at flow time zero, because the fields
$c$ and $\cbar$ always occur in pairs and only the
product of their renormalization factors matters.

\subsection 5.3 Do the bulk fields require renormalization?

To one-loop order of perturbation
theory, the question may be answered by explicitly calculating
the divergent parts (if any) of bulk-field correlation functions.
We first consider the 
two-point function of the $B$ field and
define its Lorentz-invariant parts ${\cal A}$ and ${\cal B}$
through
\equation{
   \bigl\langle\tilde{B}_{\mu}^a(t,p)\tilde{B}_{\nu}^b(s,q)\bigr\rangle
   =(2\pi)^D\delta(p+q){\delta^{ab}\over(p^2)^2}
   \noenum
   \nexteq{2.5ex}
   {\phantom{   \langle\tilde{B}_{\mu}^a(t,p)\tilde{B}_{\nu}^b(s,q)\rangle
   =}}
   \times
   \bigl\{(\delta_{\mu\nu}p^2-p_{\mu}p_{\nu}){\cal A}(t,s,p^2)
   +p_{\mu}p_{\nu}{\cal B}(t,s,p^2)\bigr\}.
   \enum
}
All self-energy diagrams drawn in fig.~4 contribute to 
${\cal A}$ and ${\cal B}$. 
The diagrams $4-6$ actually make 
two contributions, because their external legs are different.

In the renormalized perturbation expansion 
\equation{
   {\cal X}=\mu^{2\eps}\sum_{l=0}^{\infty}g^{2l+2}{\cal X}^{(l)},
   \qquad
   {\cal X}={\cal A}\;\hbox{or}\;{\cal B},
   \enum
}
the leading-order coefficients are 
\equation{
   {\cal A}^{(0)}=\rme^{-(t+s)p^2},
   \qquad
   {\cal B}^{(0)}=\lambda^{-1}\rme^{-\alpha_0(t+s)p^2}.
   \enum
}
At the next order, the residues of the poles that
derive from the renormalization of the coupling and the gauge-fixing
parameter are thus given by
\equation{
   \left.{\rm res}\{{\cal A}^{(1)}\}\right|_{Z\;{\rm factors}}=
   -b_0{\cal A}^{(0)},
   \enum
   \nexteq{2.5ex}
   \left.{\rm res}\{{\cal B}^{(1)}\}\right|_{Z\;{\rm factors}}=
   (c_0-b_0){\cal B}^{(0)}.
   \enum
}
Recalling the results obtained in subsect.~5.1, 
it is now straightforward to verify that these poles 
are canceled by the divergent parts of the one-loop diagrams.
Up to this order of perturbation theory,
the two-point function of the time-dependent gauge field
is thus finite and does not require further renormalization.

In the case of the $BL$, $B\ren{A}$, $L\ren{A}$ and $LL$ correlation 
functions, the cancellation of the singularities at $D=4$ can be
shown in the same way. There is actually little to prove in
the last two instances, because these two-point functions vanish
to all orders of perturbation theory (there are no diagrams
with ingoing and no outgoing flow lines). All calculations
reported in this section thus support the conjecture that 
the $B$ and the $L$ field do not need to be renormalized.

\section 6. BRS symmetry

We now proceed with the general discussion that will lead to 
the proof of finiteness of the correlation functions of the bulk 
fields to all orders in the gauge coupling.
As a first step,
the BRS symmetry of the theory in $D+1$ dimensions 
is reviewed in this section.

The gauge fixing discussed in sect.~3.1
follows the standard procedure and therefore leads to
a theory with a BRS symmetry.
This symmetry acts on the boundary fields in the usual way, but
the transformation of the bulk fields requires the solution of 
the diffusion equation (3.2) and consequently tends to be non-local.
In their work on the renormalization
of the Langevin equation, Zinn--Justin and Zwanziger [\ref{ZinnZwanziger}]
however showed that the locality of the transformation can be
restored by introducing additional ghost fields.

\subsection 6.1 BRS transformation of the boundary fields

The BRS variation of the unrenormalized
fields $A_{\mu}$, $\cbar$ and $c$ is defined by [\ref{BRSI},\ref{BRSII}]
\equation{
   \dbrs A_{\mu}=D_{\mu}c,
   \enum
   \nexteq{2.0ex}
   \dbrs c=-c^2,
   \enum
   \nexteq{2.0ex}
   \dbrs\cbar=\lambda_0\partial_{\mu}A_{\mu}.
   \enum
}
Note that the components $c^a$ and $\cbar^a$ of the ghost fields 
and the operator $\dbrs$ 
anti-commute with one another.
The product on the right of eq.~(6.2), for example, 
is given by
\equation{
   c^2=c^ac^bT^aT^b=\frac{1}{2}c^ac^bf^{abc}T^c,
   \enum
}
and the Leibniz rule for the operator $\dbrs$ 
must take its anti-commuting character
into account.
The action $S+\Sgf+\Sgh$ 
and the measure in the functional integral 
of the theory in $D$ dimensions are then 
easily shown to be BRS invariant.

\subsection 6.2 Bulk ghost fields

The additional ghost fields $d(t,x)$ and $\dbar(t,x)$ 
mentioned above live in $D+1$ dimensions but are otherwise
of the same kind as the Faddeev--Popov ghosts.
Their action is
\equation{
   \Sflgh=-2\int_{0}^{\infty}\rmd t\int\rmd^Dx\,
   \tr\bigl\{
   \dbar(t,x)\bigl(\partial_td-\alpha_0 D_{\mu}\partial_{\mu}d\bigr)(t,x)
   \bigr\}
   \enum
}
and the $d$ field is required to satisfy the boundary condition
\equation{
   \left.d\right|_{t=0}=c.
   \enum
}
No boundary condition is imposed on the $\dbar$ field, which, in many respects,
plays a r\^ole similar to the Lagrange-multiplier field $L_{\mu}$
in the case of the gauge field.

The propagator of the ghost fields,
\equation{
   \langle\dtilde^a(t,p)\dbartilde^b(s,q)\rangle=
   (2\pi)^D\delta(p+q)\delta^{ab}\theta(t-s)\tilde{K}_{t-s}(p)
   +\rmO(g_0^2),
   \enum
   \nexteq{3.0ex}
   \tilde{K}_{t}(p)=\rme^{-\alpha_0tp^2},
   \enum
}
can be worked out following the steps taken in subsect.~4.2.
Graphically the propagator is represented by a dashed line,
\equation{
  \raise-0.68cm\hbox{\epsfxsize=1.9cm\epsfbox{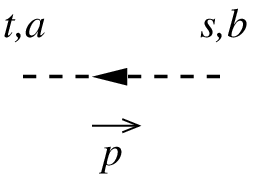}}%
  \;=\;\delta^{ab}\theta(t-s)\tilde{K}_{t}(p),
  \enum
}
where the arrow is drawn in the direction from $\dbar$ to $d$.
There is a mixed propagator,
\equation{
   \langle\dtilde^a(t,p)\cbartilde^b(q)\rangle=
   (2\pi)^D\delta(p+q)\delta^{ab}g_0^2\Dtilde_t(p)
   +\rmO(g_0^4),
   \enum
   \nexteq{3.0ex}
   \Dtilde_t(p)={1\over p^2}\rme^{-\alpha_0tp^2},
   \enum
}
as well, which coincides with the $c\cbar$ propagator at $t=0$ and 
is therefore represented by the same graphical symbol 
(a directed dotted line). 
The $c\dbar$ two-point function, on the other hand, 
vanishes to all orders.

The action
\equation{
   \left.\Sflgh\right|_{\rm interaction}=
   -\int_0^\infty\rmd t\int_{p,q,r}(2\pi)^D\delta(p+q+r)
   \noenum
   \nexteq{2.0ex}
   {\phantom{\left.\Sflgh\right|_{\rm interaction}=-}}
   \times\vbulk{1,1}(p,q,r)^{abc}_{\mu}
   \tilde{B}^a_{\mu}(t,-p)\dbartilde(t,-q)^b\dtilde(t,-r)^c,
   \enum
}
also gives rise to a new flow vertex $\vbulk{1,1}$. 
In the Feynman diagrams,
it is represented by an open circle as the other flow vertices.
The explicit expression for the vertex is given in appendix B.

As in the case of the flow-line loops discussed in subsect.~4.4, it
is possible to show that diagrams with $d\dbar$ loops vanish.
Note that ghost loops with 
mixed propagators do not exist, because there is no $c\dbar$ propagator 
and the vertices only couple
$c$ to $\cbar$ or $d$ to $\dbar$. The only non-zero 
ghost loops are thus the
usual ones of the theory at flow time zero.

\subsection 6.3 BRS variation of the fields in the bulk

The BRS symmetry acts on the bulk fields 
according to [\ref{ZinnZwanziger}]
\equation{
   \dbrs B_{\mu}=D_{\mu}d,
   \enum
   \nexteq{2.0ex}
   \dbrs L_{\mu}=[L_{\mu},d],
   \enum
   \nexteq{2.0ex}
   \dbrs d=-d^2,
   \enum
   \nexteq{2.0ex}
   \dbrs\dbar=D_{\mu}L_{\mu}-\{d,\dbar\}.
   \enum
}
Note that $\dbrs B_{\mu}=\dbrs A_{\mu}$ and 
$\dbrs d=\dbrs c$ at the boundary, as must be the case
in view of the boundary conditions (2.3) and (6.6).

In order to show that the BRS variation of the 
bulk action $\Sfl+\Sflgh$ vanishes, it is helpful to 
introduce the fields
\equation{
   E_{\mu}=\partial_tB_{\mu}-D_{\nu}G_{\nu\mu}
   -\alpha_0D_{\mu}\partial_{\nu}B_{\nu},
   \enum
   \nexteq{3.0ex}
   e=\partial_td-\alpha_0 D_{\mu}\partial_{\mu}d.
   \enum
}
After some algebra, one finds that
\equation{
   \dbrs E_{\mu}=[E_{\mu},d]+D_{\mu}e,
   \enum
   \nexteq{3.0ex}
   \dbrs e=-\{e,d\},
   \enum
}
and the invariance of the bulk action is then easily established.

\topinsert
\vbox{
\vskip0.0cm
\centerline{\epsfxsize=10.0cm\epsfbox{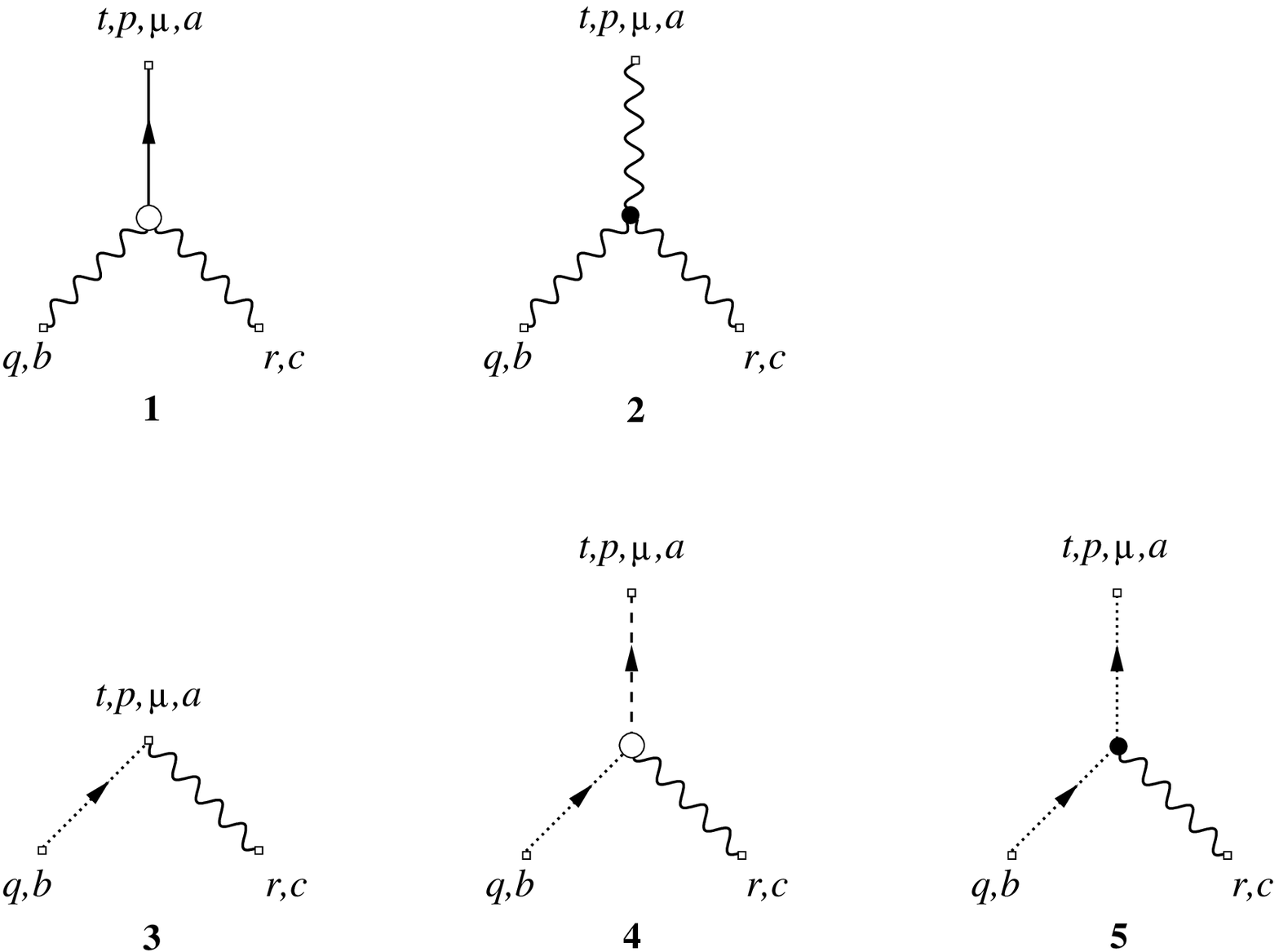}}
\vskip0.5cm
\figurecaption{%
Tree diagrams contributing to the correlation function on the left 
of eq.~(6.23) (diagrams 1 and 2) and to the one on the right
of the equation (diagrams 3,4 and 5). 
The momenta are ingoing and satisfy $p+q+r=0$. In diagram $3$,
the momentum $p$ flows into a vertex representing
the insertion of the field $[B_{\mu},d]$, while in the case
of the diagrams $4$ and $5$ the external lines with momentum $p$
are multiplied by $ip_{\mu}$ 
(thus representing the field $\partial_{\mu}d$).
}
}
\endinsert

\subsection 6.4 Examples of BRS identities

Since the functional integration measure is also invariant, it follows
that $\langle\dbrs{\cal O}\rangle=0$ for any product 
${\cal O}$ of the basic fields.
This leads to identities such as
\equation{
   \lambda_0\langle
   \partial_{\mu}A_{\mu}^a(x)\partial_{\nu}A_{\nu}^b(y)
   \rangle=g_0^2\delta^{ab}\delta(x-y),
   \enum
   \nexteq{3.0ex}
   \lambda_0\langle B^a_{\mu}(t,x)\partial_{\nu}A^b_{\nu}(y)\rangle=
   -\langle(D_{\mu}d)^a(t,x)\cbar^b(y)\rangle,
   \enum
   \nexteq{3.0ex}
   \lambda_0\langle 
   B^a_{\mu}(t,x)\partial_{\nu}A^b_{\nu}(y)\partial_{\rho}A^c_{\rho}(z)
   \rangle=
   -\langle(D_{\mu}d)^a(t,x)\cbar^b(y)\partial_{\rho}A^c_{\rho}(z)\rangle,
   \enum
}
the first of them being the familiar gauge Ward identity in the
theory in $D$ dimensions,
which determines the longitudinal part of the gauge-field 
two-point function.

It is instructive to check eqs.~(6.22) and (6.23) 
at tree level of perturbation theory. The first
equation relates the longitudinal part of the mixed 
gauge-field propagator to the mixed ghost propagator, an identity
that is immediate from the expressions for the propagators.
The other equation involves the $3$-point vertices 
(see fig.~5) and a non-trivial
cancellation among various contributions (appendix C).

\section 7. Finiteness of the renormalized perturbation expansion

The renormalization constants 
$Z$, $Z_3$ and $\tilde{Z}_3$ are now assumed to be such
that the singularities of the 
correlation functions of the fields $(\ren{A})_{\mu},\ren{c}$
and $\ren{\cbar}$ at $D=4$ 
cancel to all orders of the renormalized coupling.
Our aim in this section is to show that
the correlation functions involving the bulk fields
at positive flow time are then finite too and thus
do not require further renormalization.

\subsection 7.1 Renormalized perturbation theory

We first need to reorganize the perturbation expansion
in terms of the renormalized parameters and fields.
In this form,
the expansion is generated by an action $S_0+\Delta S$
of the renormalized fields,
which includes the counterterms $\Delta S$ required to
cancel the poles of the Feynman diagrams at $D=4$. 

Explicitly $S_0$ and $\Delta S$ are obtained by
expressing the bare parameters and fields 
in the total action
\equation{
  \Stot=S+\Sgf+\Sgh+\Sfl+\Sflgh
  \enum
}
through the renormalized ones and by setting
\equation{
  S_0=\left.\Stot\right|_{Z=Z_3=\tilde{Z}_3=1},
  \enum
  \nexteq{2.5ex}
  \Delta S=\Stot-S_0+\Delta\Sbc,
  \enum
}
where
\equation{
   \Delta\Sbc=2\int\rmd^Dx\,\tr\bigl\{
   (Z^{1/2}Z_3^{1/2}-1)L_{\mu}(0,x)(\Ar)_{\mu}(x)
   \noenum
   \nexteq{2.0ex}
   {\phantom{\Delta S_{\rm bc}=2\int\rmd^Dx\,\tr\bigl\{}}
   +
   (\tilde{Z}_3Z^{1/2}Z_3^{1/2}-1)\dbar(0,x)\ren{c}(x)\bigr\}.
   \enum
}
The boundary conditions are then
\equation{
  \left.B_{\mu}\right|_{t=0}=(\ren{A})_{\mu},
  \qquad
  \left.d\right|_{t=0}=\ren{c},
  \enum
}
and the Feynman rules derived from the action $S_0$ 
thus coincide with those discussed in sect.~3
(apart from the fact that the bare parameters $g_0$ and $\lambda_0$
are replaced by $\mu^{\eps}g$ and $\lambda$, respectively).

Note that the boundary conditions (7.5) differ from
the ones imposed on the bare fields (a product of 
renormalization factors is missing).
The counterterm $\Delta\Sbc$ must be included
in the action of the renormalized fields
to correct for this.
It amounts to adding two-point vertices to the 
Feynman rules, whose effect on the 
gauge-field and ghost propagators is 
equivalent to a change of the boundary 
conditions by the missing renormalization factors.

\subsection 7.2 Absence of bulk counterterms

In the renormalized perturbation expansion,
singularities at $D=4$ (if any) appear for the first time
at some loop order $l\geq1$.
Since the interactions are local, and since
all propagators are tempered distributions in position space 
with singularities only at coinciding arguments, 
the divergent parts of the correlation functions at 
this loop order are expected to be
such that they can be canceled by local counterterms.
Similarly to the case of an ordinary
field theory on a half-space studied by Symanzik [\ref{Symanzik}],
the counterterms can be localized either in the bulk or at
the boundary of the half-space. 

In the theory considered here,
bulk counterterms can be excluded from the outset.
In order to show this, first note that 
the correlation functions of the bulk fields
$B_{\mu}$, $L_{\mu}$, $d$ and $\dbar$ 
are, at large flow times, given by 
Feynman diagrams built from
flow lines 
and flow vertices only.
All diagrams
of this kind are directed trees or products such trees,
each tree ending at one of the $B$ and $d$ fields in
the correlation function considered. Moreover, the
flow lines at the other ends of the trees must
start from the $L$ and $\dbar$ fields in the
correlation function.

Since there are no loop diagrams,
the correlation functions of the bulk fields are
non-singular at large flow times and do not require
renormalization.
Divergent bulk counterterms are therefore excluded.
Note, incidentally, that
correlation functions of local fields composed from $B$ and $d$ 
fields are finite too, because no loop
diagrams are generated when the arguments of some of 
these fields coalesce.
The field $D_{\mu}d$ that appears in the BRS 
identity (6.23), for example, is of this kind
and consequently does not need to be renormalized.

\subsection 7.3 Boundary counterterms

The discussion in the previous
subsection shows that the structure of any divergent
parts of the correlation functions 
must correspond to the insertion
of a counterterm localized at flow time zero.
Since the correlation functions of the renormalized fields 
$(\ren{A})_{\mu},\ren{c}$ and $\ren{\cbar}$ 
are finite to all orders, divergences can only
arise from diagrams with at least one flow vertex
and thus at least one such vertex with an external flow line.
The possible counterterms are therefore proportional to 
$L_{\mu}$ or $\dbar$.

Since the gauge coupling is dimensionless, 
the counterterm
may not involve composite fields of dimension larger than $4$ 
(such terms would be irrelevant). Moreover, Lorentz symmetry,
global gauge invariance and the ghost number conservation 
must be respected.
Since $L_{\mu}$ and $\dbar$ have dimension $3$
and all other fields dimension $1$, it follows that
the possible counterterms at $l$-loop order are of the form
\equation{
   2g^{2l}\int\rmd^Dx\,\tr\bigl\{
   z_1L_{\mu}(0,x)(\Ar)_{\mu}(x)+
   z_2\dbar(0,x)\ren{c}(x)\bigr\}
   \enum
}
with some (singular) coefficients $z_1$ and $z_2$.
An $LB$ and a $\dbar d$ term should in principle be 
included here, but in view of the boundary conditions (7.5),
these terms are not independent and their inclusion
would be equivalent to a change of 
the coefficients $z_1$ and $z_2$.

\subsection 7.4 Consequences of the BRS symmetry

We now show that the boundary counterterm (7.6) is excluded
by the BRS symmetry of the theory.
In terms of the renormalized fields and parameters, the 
BRS identity (6.23) reads
\equation{
   \lambda\langle 
   B^a_{\mu}(t,x)\partial_{\nu}(\ren{A})^b_{\nu}(y)
   \partial_{\rho}(\ren{A})^c_{\rho}(z)
   \rangle=
   -\langle(D_{\mu}d)^a(t,x)
   (\ren{\cbar})^b(y)\partial_{\rho}(\ren{A})^c_{\rho}(z)\rangle.
   \enum
}
The unusual asymmetric renormalization (5.13) of the ghost fields 
$c$ and $\cbar$ was chosen to ensure that the renormalization factors 
drop out in this equation. From the point of view of the
$\SUn$ gauge theory, the standard and 
the asymmetric renormalization of the ghost fields are equivalent,
but the situation
is different in the theory in $D+1$ dimensions, because only $c$ couples
to the bulk fields.

Equation (7.7) holds as long as no extra counterterms need to be
added and thus up to loop order $l$ inclusive. Note that
all fields in the correlation functions in this
equation, including the composite field $D_{\mu}d$,
are renormalized fields that cannot be additionally
renormalized (cf.~subsect.~7.2).
If the correlation functions
are singular at $l$-loop order, one must therefore
be able cancel the singularities by a counterterm of 
the form (7.6).

The contribution of the counterterm (7.6) to the correlation 
functions is obtained by inserting the corresponding two-point
vertices in the tree diagrams in fig.~5. The insertions have
the effect of multiplying the values of the diagrams $1-5$ by
$g^{2l}$ times
\equation{
   2z_1,\,z_1,\,z_1+z_2,\,z_1+z_2\;\hbox{and}\;z_2
   \enum
}
respectively. However,
recalling the values of the diagrams 
quoted in appendix C, the sum of the diagrams 
weighted by the factors (7.8) turns out to violate 
the BRS identity (7.7) unless $z_1=z_2=0$. 

The correlation functions in eq.~(7.7) must therefore be finite
at $l$-loop order and all other (renormalized) correlation functions 
must be non-singular too, because the addition to the action 
of a counterterm of the form (7.6) is excluded.
We have thus shown that the theory in $D+1$ 
dimensions does not require further renormalization.

\section 8. Miscellaneous remarks

\vskip-2.5ex

\subsection 8.1 Behaviour of the gauge field near flow time zero

In the regularized theory, the time-dependent field $B_{\mu}(t,x)$
satisfies the boundary condition (2.3) and its correlation functions
thus converge to those of the bare gauge field $A_{\mu}(x)$ 
when $t$ goes to zero. However,
since the latter requires renormalization by a divergent constant,
the correlation functions tend to become singular at $t=0$
after renormalization and removal of the regularization.
Their asymptotic behaviour for $t\to0$ is then
described by an expansion
\equation{
   B_{\mu}(t,x)=c_B(t)(\ren{A})_{\mu}(x)+\rmO(t)
   \enum
}
in local renormalized fields with singular coefficients [\ref{Symanzik}]. 
At one-loop order of perturbation theory, for example,
the coefficient $c_B(t)$ is found to 
diverge logarithmically at $t=0$.

It is straightforward to show that the
renormalization group equation
\equation{
   \left\{\mu{\partial\over\partial\mu}+\beta{\partial\over\partial g}
   -2\gamma\lambda{\partial\over\partial\lambda}
   +\gamma+\beta/g
   \right\}c_B(t)=\rmO(t)
   \enum
}
holds, where
\equation{
   \beta=-b_0g^3+\rmO(g^5),
   \qquad
   \gamma=c_0g^2+\rmO(g^4),
   \enum
}
are the beta function and the anomalous dimension of the
gauge field. In the Landau gauge, for example, 
the equation can be easily integrated and one finds that
\equation{
   B_{\mu}(t,x)
   \mathrel{\mathop\sim_{t\to0}}
   \bigl(2b_0\gbar(q)^2\bigr)^{1/2-c_0/2b_0}
   R(g)(\ren{A})_{\mu}(x),
   \enum
}
$\gbar(q)$ being the running coupling at momentum $q=(8t)^{-1/2}$
and $R(g)$ the factor that relates the renormalized to 
the renormalization-group-invariant gauge field.

Equation (8.4) is a remnant of the
boundary condition satisfied by the gauge field in the regularized theory.
It shows that the field generated by the flow equation
is connected to the fundamental field in a universal manner, i.e.~there
is no room for finite renormalizations here, the reason being
that any such renormalization would violate the BRS symmetry.

\subsection 8.2 Gauge-invariant composite fields

Wilson loops and gauge-invariant local fields are independent of 
the parameter $\alpha_0$ in the flow equation 
and do not require renormalization 
at positive flow time. At small flow times,
their asymptotic behaviour is determined by the scaling 
properties of the corresponding
renormalized fields in the $\SUn$ gauge theory and thus
reflects the singular nature of the latter.

For illustration, consider the density
\equation{
   E=\frac{1}{4}G_{\mu\nu}^aG_{\mu\nu}^a
   \enum
}
that has previously been studied in ref.~[\ref{WilsonFlow}].
Since $E$ can mix with the unit field, the dominant term
at small flow times,
\equation{
   E(t,x)=\langle E(t,x)\rangle+c_E(t)
   \ren{\bigl\{\frac{1}{4}F^a_{\mu\nu}F^a_{\mu\nu}\bigr\}}(x)+\rmO(t),
   \enum
}
is its expectation value, while the first subleading term is proportional
to the renormalized action density of the fundamental field.
The associated coefficients, 
$\langle E(t,x)\rangle$ and $c_E(t)$,
satisfy a renormalization 
group equation that allows their exact asymptotic behaviour
at small $t$ to be worked out analytically.

\subsection 8.3 Theories with matter fields

In the presence of matter fields, the gradient flow is
defined by the same equations 
as in the pure gauge theory (subsect.~2.1).
The interesting but rather more complicated case where
the matter fields are included in the time evolution
is not considered here.

In a renormalizable theory, matter fields are
scalar fields of dimension $1$ or fermion fields of 
dimension $3/2$. Provided
the regularization preserves the
gauge symmetry, our argumentation 
in sects.~6 and 7 then carries over literally. In particular,
additional boundary counterterms involving the matter fields
are excluded by the global symmetries and the requirement
that their dimension must be less than or equal to $4$. 
The finiteness of the correlation functions
at positive flow times is therefore again guaranteed.

QCD is a prominent example of a gauge theory that 
has all the required properties to ensure finiteness,
but the same applies to many more theories of interest, 
including the SU(2) Higgs model, 
supersymmetric versions of QCD and technicolour theories.

\subsection 8.4 Lattice regularization

While the Feynman rules are more 
complicated than with dimensional regularization,
the perturbative analysis of the gradient flow on the lattice 
is not expected to run into fundamental difficulties.

A possible choice of the gradient term
in the lattice flow equation
is the gradient of the Wilson action
[\ref{WilsonFlow}].
There is however no reason to choose this particular action
or to require that it coincides with the gauge action 
of the theory.
As long as the term has the 
correct form and normalization in the classical continuum limit
(and thus at tree-level of perturbation theory), 
the correlation functions of the time-dependent gauge field
will not depend on the exact choices one makes, except 
for lattice effects vanishing proportionally to a positive power
of the lattice spacing $a$.

The continuous Lorentz symmetry is broken on the lattice,
but the remaining exact symmetries are sufficient to exclude
counterterms that have not already appeared
in the continuum theory.
We therefore expect that the correlation functions
of the time-dependent fields
do not require renormalization and
that their values in the continuum limit 
are independent of the regularization
(up to finite renormalizations of 
the coupling and the gauge-fixing parameter).
Moreover, if the lattice theory is $\rmO(a)$ improved,
the correlation functions are automatically improved too,
because there are no candidate $\rmO(a)$ counterterms 
with all the required properties.

\section 9. Conclusions

The fact that the gauge field generated by the gradient flow 
does not require renormalization is a consequence of the
locality, the symmetries and the parabolic nature
of the flow equation.
In particular, in the
associated field theory in $D+1$ dimensions,
bulk counterterms are excluded simply
because the evolution of the bulk fields is retarded at large times
and thus described by tree diagrams.
The absence of 
boundary counterterms other than those needed for the renormalization of 
the theory at flow time zero is however
non-trivial and can only be shown 
using power-counting and the BRS symmetry.

In presence of matter fields, the situation is essentially 
unchanged as long as only the gauge field is evolved in time.
The finiteness of the correlation functions of the field at positive
flow time is therefore still guaranteed.
Including all or some of the matter fields in the flow is however
an interesting option that remains to be explored.

\vskip1.0ex
We thank Jean Zinn--Justin for helpful discussions on
the renormalization of the Langevin equation and for encouraging us
to proceed directly with the flow equation rather than considering
the latter to be a limit of the Langevin equation.

\appendix A. Notational conventions

The Lie algebra $\sun$ of $\SUn$ 
may be identified with the linear space of all anti-hermitian
traceless $N\times N$ matrices.
With respect to a 
basis $T^a$, $a=1,\ldots,N^2-1$, of such matrices, 
the elements $X\in\sun$ are given by
$X=X^aT^a$ with real components $X^a$
(repeated group indices are automatically summed over).
The structure constants $f^{abc}$ in
the commutator relation
\equation{
  [T^a,T^b]=f^{abc}T^c
  \enum
}
are real and totally anti-symmetric in the indices if
the normalization condition
\equation{
  \tr\{T^aT^b\}=-\frac{1}{2}\delta^{ab}
  \enum
}
is imposed.
Moreover, $f^{acd}f^{bcd}=N\delta^{ab}$.

Gauge fields in the continuum theory 
take values in the Lie algebra of the gauge group.
Lorentz indices $\mu,\nu,\ldots$ are automatically
summed over when they occur in matching pairs.
The space-time metric is assumed to be euclidean. In particular,
$p^2=p_{\mu}p_{\mu}$ for any momentum $p$ and $\delta_{\mu\mu}=D$.

\appendix B. Flow vertices

The flow vertices $\vbulk{2,0},\vbulk{3,0}$ and $\vbulk{1,1}$ 
are defined through eqs.~(2.13) and (6.12), respectively.
They are explicitly given by
\equation{
  \vbulk{2,0}(p,q,r)^{abc}_{\mu\nu\rho}=
  if^{abc}\bigl\{(r-q)_{\mu}\delta_{\nu\rho}
  +2q_{\rho}\delta_{\mu\nu}-2r_{\nu}\delta_{\mu\rho}
  \noenum
  \nexteq{2.5ex}
  {\phantom{\vbulk{2,0}(p,q,r)^{abc}_{\mu\nu\rho}=
  if^{abc}\bigl\{}}
  +(\alpha_0-1)
  (q_{\nu}\delta_{\mu\rho}-r_{\rho}\delta_{\mu\nu})\bigr\},
  \enum
  \nexteq{3.5ex}
  \vbulk{3,0}(p,q,r,s)^{abcd}_{\mu\nu\rho\sigma}=
  f^{abe}f^{cde}
  (\delta_{\mu\sigma}\delta_{\nu\rho}-\delta_{\mu\rho}\delta_{\sigma\nu})
  \noenum
  \nexteq{2.5ex}
  \qquad\quad
  +f^{ade}f^{bce}
  (\delta_{\mu\rho}\delta_{\sigma\nu}-\delta_{\mu\nu}\delta_{\rho\sigma})+
  f^{ace}f^{dbe}
  (\delta_{\mu\nu}\delta_{\rho\sigma}-\delta_{\mu\sigma}\delta_{\nu\rho}),
  \enum
  \nexteq{3.5ex}
  \vbulk{1,1}(p,q,r)^{abc}_{\mu}=\alpha_0if^{abc}r_{\mu}.
  \enum
} 
In the Feynman diagrams, the momenta in these formulae are identified
with the ingoing line momenta.

\appendix C. Evaluation of the diagrams in fig.~5

The contributions of the diagrams in fig.~5 to the 
Fourier transform of the correlation functions
in the BRS identity (6.23) are of the form
\equation{
   -{g_0^4if^{abc}\over \lambda_0 p^2q^2r^2}{\cal C}_{\mu,n},
   \enum
}
where $n$ is the diagram number. Explicitly, one obtains
\equation{
   {\cal C}_{\mu,1}=
   \frac{1}{2}p^2(q-r)_{\mu}\rme^{-t\alpha_0(q^2+r^2)}+
   \frac{1}{2}(q^2-r^2)p_{\mu}\rme^{-t\alpha_0p^2}
   -{\cal C}_{\mu,2},
   \enum
   \nexteq{3.0ex}
   {\cal C}_{\mu,2}=
   \frac{1}{2}\bigl\{p^2(q-r)_{\mu}+(q^2-r^2)p_{\mu}\bigr\}\rme^{-tp^2},
   \enum
}
for the diagrams contributing to the left-hand side of the equation
and
\equation{
   {\cal C}_{\mu,3}=-p^2r_{\mu}\rme^{-t\alpha_0(q^2+r^2)},
   \enum
   \nexteq{3.0ex}
   {\cal C}_{\mu,4}=-\frac{1}{2}p^2p_{\mu}\bigl\{
   \rme^{-t\alpha_0(q^2+r^2)}-\rme^{-t\alpha_0p^2}\bigr\},
   \enum
   \nexteq{3.0ex}
   {\cal C}_{\mu,5}=-(qr+r^2)p_{\mu}\rme^{-t\alpha_0p^2},
   \enum
}
for the diagrams on the right-hand side. The BRS symmetry requires
\equation{
   {\cal C}_{\mu,1}+{\cal C}_{\mu,2}=
   {\cal C}_{\mu,3}+{\cal C}_{\mu,4}+{\cal C}_{\mu,5},
   \enum
}
which is indeed the case.

\beginbibliography


\bibitem{WilsonFlow}
M. L\"uscher,
{\it Properties and uses of the Wilson flow in lattice QCD},
JHEP 1008 (2010) 071

\bibitem{Villasimius}
M. L\"uscher,
{\it Topology, the Wilson flow and the HMC algorithm},
XXVIII International Symposium on Lattice Field Theory, June 14-19 2010, Villasimius, Italy, 
PoS(Lattice 2010)015


\bibitem{Zinn}
J. Zinn--Justin,
{\it Renormalization and stochastic quantization},
Nucl. Phys. B275 [FS17] (1986) 135

\bibitem{ZinnZwanziger}
J. Zinn--Justin, D. Zwanziger, 
{\it Ward identities for the stochastic quantization of gauge fields},
Nucl. Phys. B295 [FS21] (1988) 297


\bibitem{BRSI}
C. Becchi, A. Rouet and R. Stora,
{\it Renormalization of the abelian Higgs-Kibble model},
Comm. Math. Phys. 42 (1975) 127

\bibitem{BRSII}
C. Becchi, A. Rouet and R. Stora,
{\it Renormalization of gauge theories},
Ann. Phys. (NY) 98 (1976) 287


\bibitem{Symanzik}
K. Symanzik,
{\it Schr\"odinger representation and Casimir effect in 
renormalizable quantum field theory},
Nucl. Phys. B190 [FS3] (1981) 1

\endbibliography

\bye